\documentstyle[prl,epsf,floats,aps,amsfonts,twocolumn]{revtex}

\begin{document}

\twocolumn[\hsize\textwidth\columnwidth\hsize\csname
@twocolumnfalse\endcsname

\title{Free expansion of lowest Landau level states of trapped
atoms:\\
a wavefunction microscope}
\author{N. Read$^1$ and N.R. Cooper$^2$}
\address{$^1$ Department of Physics, Yale
University, P.O. Box 208120, New Haven, CT 06520-8120\\
$^2$Theory of Condensed Matter Group, Cavendish Laboratory,
Madingley Road, Cambridge, CB3 0HE, United Kingdom.}
\date{June 13, 2003}
\maketitle

\begin{abstract}
We show that for any lowest-Landau-level state of a trapped,
rotating, interacting Bose gas, the particle distribution in
coordinate space in a free expansion (time of flight) experiment
is related to that in the trap at the time it is turned off by a
simple rescaling and rotation. When the lowest-Landau-level
approximation is valid, interactions can be neglected during the
expansion, even when they play an essential role in the ground
state when the trap is present. The correlations in the density in
a single snapshot can be used to obtain information about the
fluid, such as whether a transition to a quantum Hall state has
occurred.
\end{abstract}

%\pacs{PACS numbers: }
]

%%%%%%%%%%%%%%%%%%%%%%%%%%%%%%%%%%%%%%%%%%%%%%%%%%%%%%%%%%%

There has recently been interest in states of trapped atoms that
are rotating near the critical frequency of a harmonic trap
\cite{wgs,gp,yrast,wgcw,cwg,qh,hm,shm1,shm2}. When the atoms are
bosons, there is a vortex lattice at moderate rotation rates
\cite{gp}, but at high rotation rates quantum fluctuations are
predicted to melt the lattice\cite{cwg,shm1} and produce exotic
highly-correlated many-body states that are related to those
occurring for electrons in the quantum Hall effect
\cite{wgs,wgcw,cwg,qh,hm}. A standard experimental technique is to
turn off the trap potential suddenly, and then take a snapshot of
the cloud of atoms after it has expanded to many times its
original size (a free-expansion or time-of-flight experiment). A
question that arises is what information can be extracted from
such an image of a quantum Hall state. In this note, we point out
that in a relevant regime, the density distribution in such a
snapshot directly represents the density at the time of
switch-off, but rescaled and rotated by 90$^\circ$ about the
original rotation axis. This result is simpler than the results of
previous work on the free expansion of a Bose condensate in the
Gross-Pitaevskii or Thomas-Fermi regimes, that can describe a Bose
condensate or vortex lattice. While in a few cases the length
scales of the density distribution are simply scaled up by a
factor, more generally the condensate function also evolves during
expansion \cite{kss,cd1,cd2,dm}. Usually, the form of the density
distribution after a free expansion is not simply related to the
initial density; rather, for noninteracting particles it is
related to the momentum distribution. Although our result can be
extracted from special cases of earlier analyses, the emphasis is
different. We emphasize that the neglect of interactions during
the free expansion is justified whenever all the bosons can be
assumed to be in the lowest Landau level (LLL) just before the
switch-off, even though interactions may be crucial in the
highly-correlated ground state that exists before that time.
Consequently, the result applies to any LLL many-body state, not
only a Bose condensate. (These points were also made briefly in
Ref.\ \cite{hm} for the density expectation value.) We also
suggest how a snapshot of the particles after a free expansion can
be used to obtain information about the nature of the original
state, as it constitutes a microscope that enlarges the real-space
image of the state.

We begin by explaining the classical version of the problem, which
will make the later quantum many-body treatment easily
understandable. We take particles of mass $M$ in a harmonic trap,
so the particles move in a three-dimensional harmonic oscillator
potential, with frequency $\omega_3$ for oscillation in the $3$
direction, and $\omega_\perp$ for oscillation in the $1$--$2$
plane. Thus in general our model has rotational symmetry about the
$3$ axis only. Consider a single particle in this potential. The
general form of an orbit is an ellipse centered at the center of
the trap. However, we wish to focus on the single-particle states
referred to in the quantum case as the LLL. These are the states
of lowest kinetic energy for each positive value of the angular
momentum about the $3$ axis. The corresponding classical orbits
are circles in the $1$--$2$ plane, circling the $3$ axis in the
positive direction. For such an orbit, the momentum when the
particle is at ${\bf r}=(x_1,x_2,x_3)$ is clearly ${\bf
p}=M\omega_\perp \hat{\bf n}_3\times{\bf r}$, where $\hat{\bf
n}_3$ is a unit vector in the $3$ direction, and the angular
momentum is ${\bf r}\times {\bf p}=M\omega_\perp |{\bf r}|^2
\hat{\bf n}_3$. If the trap potential is removed at time $t=0$, at
which ${\bf r}={\bf r}_0$, ${\bf p}={\bf p}_0$, the particle
travels freely in a straight line. At time $t$ it will be at ${\bf
r}_0+{\bf p}_0t/M={\bf r}_0+\omega_\perp t \,\hat{\bf
n}_3\times{\bf r}_0\simeq \omega_\perp t \,\hat{\bf n}_3\times{\bf
r}_0$ for large $t$. Hence if we have a collection of particles
all moving on such orbits, their distribution in the $1$--$2$
plane a long time $t\gg 1/\omega_\perp$ after removing the trap
potential will be the same as it was at $t=0$, except for a
rescaling by $\omega_\perp t$, and a rotation by $\pi/2$ about the
$3$ axis. In this argument we have neglected the interactions
between the particles once the trap potential is turned off. We
will return to this question after discussing the quantum case.

We now turn to the fully quantum-mechanical treatment, beginning
again with a single particle. Suppose the state at time $t=0$ is
$\psi({\bf r},0)=\psi_0({\bf r})$. For the free motion, the
solution is simple in momentum space. We define the Fourier
representation of the state $\psi$ by
\begin{equation}
\psi({\bf r},t)=\int \frac{d^3k}{(2\pi)^3} \,\tilde{\psi}({\bf
k},t) e^{i{\bf k}\cdot{\bf r}},
\end{equation}
and similarly for $\psi_0$. Then the state at time $t$ is
\begin{equation}
\psi({\bf r},t)=\int \frac{d^3k}{(2\pi)^3} \,  \tilde{\psi}_0({\bf
k})e^{i{\bf k}\cdot{\bf r}-i\hbar{\bf k}^2t/(2M)}.\label{tevol}
\end{equation}

The normalized single-particle basis states for the LLL have
wavefunctions
\begin{equation}
u_m(z,x_3)=\frac{z^m e^{-|z|^2/(2\l_\perp^2) -
x_3^2/(2\l_3^2)}}{\pi^{3/4}l_\perp^{m+1}l_3^{1/2}\sqrt{m!}}.
\end{equation}
Here $z=x_1+ix_2$, the angular momentum quantum number $m=0$, $1$,
$2$, \ldots, and $l_\perp=\sqrt{\hbar/M\omega_\perp}$,
$l_3=\sqrt{\hbar/M\omega_3}$. We note that these states have their
maximum amplitude on the circles in the $1$--$2$ plane of radius
$|z|=l_\perp\sqrt{m}$, which is the same as that of the circular
classical orbit with angular momentum $m\hbar$. To calculate the
Fourier transforms, it is useful to introduce the generating
function
\begin{equation}
h_0(z,x_3)=e^{\xi z/l_\perp-|z|^2/(2\l_\perp^2) -x_3^2/(2\l_3^2)},
%/(l_\perp\pi^{3/4}l_3^{1/2}),
\end{equation}
of which the $m$th derivative with respect to $\xi$ is
proportional to $u_m(z,x_3)$. The Fourier transform of $h_0$ is
\begin{equation}
(2\pi)^{3/2} l_\perp^2 l_3 e^{-ik\xi
l_\perp-|k|^2\l_\perp^2/2-k_3^2l_3^2/2},\label{ftgen}
\end{equation}
where $k=k_1+ik_2$. Using (\ref{tevol}), at time $t$, the
generating function evolves to%
\begin{eqnarray}
\lefteqn{h(z,x_3,t)=}\,\,&&\nonumber\\&&\frac{e^{\xi z
/[l_\perp(1+i\omega_\perp t)]-|z|^2/[2l_\perp^2(1+i\omega_\perp
t)]-x_3^2/[2l_3^2(1+i\omega_3 t)]}}{(1+i\omega_\perp
t)(1+i\omega_3t)^{1/2}}.
\end{eqnarray}
It follows that if a general normalized LLL initial state is
written as
$\psi_0(z,x_3)=f(z)e^{-|z|^2/(2l_\perp^2)-x_3^2/(2l_3^2)}$, where
$f(z)$ is complex analytic in $z$ (for example, $f$ can be a
polynomial), then at time $t$ it becomes
\begin{eqnarray}
\lefteqn{\psi(z,x_3,t)=}\qquad\nonumber\\
&&\frac{f(\frac{z}{1+i\omega_\perp
t})e^{-|z|^2/[2l_\perp^2(1+i\omega_\perp
t)]-x_3^2/[2l_3^2(1+i\omega_3 t)]}}{(1+i\omega_\perp
t)(1+i\omega_3 t)^{1/2}}.%\nonumber\\&&\quad{}\times .
\end{eqnarray}
The probability density is therefore
\begin{eqnarray}
\lefteqn{|\psi(z,x_3,t)|^2=}\qquad\nonumber\\
&&\frac{|f(\frac{z}{1+i\omega_\perp
t})|^2e^{-|z|^2/[l_\perp^2(1+\omega_\perp^2
t^2)]-x_3^2/[l_3^2(1+\omega_3^2 t^2)]}}{|1+i\omega_\perp
t)(1+i\omega_3 t)^{1/2}|^2}\nonumber\\
&&=\left|\frac{\psi_0(\frac{z}{1+i\omega_\perp
t},\frac{x_3}{\sqrt{1+\omega_3^2 t^2}})}{(1+i\omega_\perp
t)(1+i\omega_3 t)^{1/2}}\right|^2. \label{genprobdens}
\end{eqnarray}
This is simply the initial probability density $|\psi_0|^2$, with
a rescaling and a rotation in the $1$--$2$ coordinates, and $x_3$
also rescaled. Notice that the factor $1+i\omega_\perp t$
describes the same rescaling and rotation of the initial complex
position $z_0=z/(1+i\omega_\perp t)$ in the $1$--$2$ plane as the
Cartesian formula ${\bf r}={\bf r}_0+\omega_\perp t \,\hat{\bf
n}_3\times{\bf r}_0$ that appeared in the classical argument. We
should point out that our result is a special limiting case of
expressions obtained by other methods \cite{kss,cd2}, on ignoring
interactions and restricting to the LLL, and is implicit in
\cite{hm}.

For large $t$ ($t\gg 1/\omega_\perp$, $1/\omega_3$), we will be
interested in $\bf r$ of order $t$, so define ${\bf V}={\bf r}/t$.
As $t \to\infty$ with ${\bf V}$ fixed, $\psi({\bf r},t)$ itself is
asymptotic to
\begin{eqnarray}
\lefteqn{\frac{f(\frac{z}{i\omega_\perp t})}{i\omega_\perp
t(i\omega_3 t)^{1/2}}e^{-|z|^2[(i\omega_\perp
t)^{-1}+(\omega_\perp t)^{-2}]
/(2l_\perp^2)}}\qquad\nonumber\\
&&{}\times e^{-x_3^2[(i\omega_3 t)^{-1}+(\omega_3
t)^{-2}]/(2l_3^2)}.
\end{eqnarray}
In this limit, the probability density (\ref{genprobdens}) becomes
\begin{equation}
\frac{\left|\psi_0\left(\frac{-iz}{\omega_\perp
t},\frac{x_3}{\omega_3 t}\right)\right|^2}{(\omega_\perp
t)^2(\omega_3 t)}.\label{probdens}
%\frac{1}{(\omega_\perp t)^2\omega_3
%t}\left|f\left(\frac{-iz}{\omega_\perp
%t}\right)e^{-|z|^2/[2l_\perp^2(\omega_\perp
%t)^2]-x_3^2/[2l_3^2(\omega_3 t)^2]}\right|^2.
\end{equation}
The transformations have simplified to a rescaling by
$\omega_\perp t$ and a rotation by $\pi/2$ in the $1$--$2$ plane,
and a rescaling by $\omega_3 t$ in the $3$ coordinate.

The long-time result can also be obtained by an application of the
stationary phase approximation to eq.\ (\ref{tevol}), which gives
the limit $t\to\infty$ with ${\bf V}={\bf r}/t$ fixed, which for a
general state $\psi_0$ is
\begin{equation}
\psi({\bf r},t) \sim \tilde{\psi}_0\left(\frac{M{\bf r}}{\hbar
t}\right)e^{iM{\bf r}^2/(2\hbar t)-3\pi
i/4}\left(\frac{M}{2\pi\hbar t}\right)^{3/2},
\end{equation}
and for the probability density
\begin{equation}
|\psi({\bf r},t)|^2 \sim \left|\tilde{\psi}_0\left(\frac{M{\bf
r}}{\hbar t}\right)\right|^2\left(\frac{M}{2\pi\hbar
t}\right)^{3}.
\end{equation}
These well-known expressions mean that at long times, the
distribution in position space is determined by the initial
momentum distribution, as if the particle propagated classically
with velocity ${\bf V}= \hbar{\bf k}/M$. This result is familiar
in optics as the Fraunhofer limit of diffraction, for example in a
two-slit experiment. In the LLL case, using eq.\ (\ref{ftgen}),
the Fourier transform of the LLL initial state is
\begin{eqnarray}
\tilde{\psi}_0({\bf k})&=&(2\pi)^{3/2}l_\perp^2l_3f(-ikl_\perp^2)
e^{-|k|^2l_\perp^2/2-k_3^2l_3^2/2}\nonumber\\
&=&(2\pi)^{3/2}l_\perp^2l_3\,\psi_0(-ikl_\perp^2,k_3l_3^2).
\end{eqnarray}
This shows that for the LLL, the probability density in $\bf k$
space is the same as that in position space, rescaled and rotated
by $\pi/2$. This corresponds both to the classical argument above,
and to the fact that in the LLL, the $1$--$2$ coordinates are
canonically conjugate, so $x_1= p_2 l_\perp^2/\hbar$, $x_2= -p_1
l_\perp^2/\hbar$. Using this in the stationary phase formula, we
recover eq.\ (\ref{probdens}).

Both the exact and approximate treatments generalize trivially to
the case of $N$ non-interacting particles with an initial state
$\psi_0({\bf r}_1, \ldots, {\bf r}_N)$, when it is a linear
combination of products of LLL single-particle states. This gives
the joint probability density for all $N$ particles after the free
expansion. This result also holds when the initial state is
described by a density matrix within the LLL, such as a state of
thermal equilibrium, but we will not consider that in detail here.

As a first example, consider a Bose condensate within the LLL. Its
wavefunction at time $t=0$ is a product,
\begin{equation}
\prod_{j=1}^N \psi_0(z_j,x_{3j}) \label{bec} \end{equation} for
some function $\psi_0$. In the LLL, if $f$ in $\psi_0$ is a
polynomial, one can always factorize
$\psi_0(z,x_3)\propto\prod_a(z-w_a)e^{-|z|^2/(2\l_\perp^2) -
x_3^2/(2\l_3^2)}$, and then we have $N_v$ vortices at complex
positions $w_a$, $a=1$, \ldots, $N_v$. After a long free
expansion, the wavefunction has the same product form, with the
bosons condensed in
\begin{eqnarray}
\lefteqn{\prod_a[z-i(\omega_\perp t)w_a]e^{-|z|^2[(i\omega_\perp
t)^{-1}+(\omega_\perp t)^{-2}]
/(2l_\perp^2)}}\qquad&&\nonumber\\&&{}\times e^{-x_3^2[(i\omega_3
t)^{-1}+(\omega_3 t)^{-2}]/(2l_3^2)}.
\end{eqnarray}
Thus the vortices will be clearly visible in an image of the
density in position space, provided the number of particles per
vortex is large (see the discussion of such images below). This is
the condition of large filling factor \cite{cwg}, which is the
regime in which a condensed ground state of the type (\ref{bec})
occurs \cite{cwg,shm1}.

As a second example, we consider the Laughlin state \cite{laugh},
$\prod_{r<s}(z_r-z_s)^2\cdot\prod_r e^{-|z_r|^2/(2\l_\perp^2) -
x_{3r}^2/(2\l_3^2)}$, where $r$, $s=1$, \ldots, $N$. The
probability density after a long time $t$ is the same as at $t=0$,
up to the usual two rescalings (the rotation can be ignored since
the Laughlin state is rotationally invariant). This implies that
all the spatial correlations are preserved after free expansion
for time $t$.

In our treatment we have completely neglected interactions during
the expansion. As justification for this, we note that, when the
trap potential is present, the LLL approximation should be valid
when the $s$-wave scattering length $a$ and typical number density
$\bar n$ in the bulk of the fluid satisfy
$4\pi\hbar^2a\bar{n}/M<2\hbar\omega_\perp$, $\hbar\omega_3$. This
condition ensures that corrections due to quantum-mechanical
mixing of non-LLL states into the ground state can be neglected.
Because $\bar{n}$ is reduced by centrifugal effects as the
rotation rate increases, this condition is more likely to be
satisfied at larger values of the total angular momenta than at
small. (For quantum Hall states, the filling factor $\nu=
\pi^{3/2}l_\perp^2l_3\bar{n}$ must be of order around $10$ or less
\cite{cwg,shm1}.) Even though the interactions are weak compared
with the kinetic energies in the trap (which are typically many
times $\hbar \omega_\perp$), the ground states may be highly
correlated. That is because, if interactions are neglected, there
are many degenerate states of the same total angular momentum
\cite{wgs}. (Classically, the particles in the ``LLL'' orbits keep
the same relative positions, up to a rotation, as they move, and
so the interactions are resonant---over a long time they can have
a large effect.)
%Alternatively, this can be understood by
%passing to the rotating frame, which is rotating at angular
%frequency $\omega$ close to $\omega_\perp$ when the LLL
%approximation is valid. In this frame, all the LLL states are
%nearly degenerate, and the interaction energy becomes very
%important in the correlated ground states \cite{cwg}.
The interaction energy scale is $4\pi\hbar^2a\bar{n}/M$, and when
the trap potential is removed, $\bar{n}$ decreases in the same way
as the probability density, eq.\ (\ref{genprobdens}). Hence the
interaction strength goes to zero after a time of order
$\max(1/\omega_\perp,1/\omega_3)$, and the correction to the
wavefunction, $\delta\psi/\psi$, due to the interaction term a
long time after switch-off is small compared with 1 if
$4\pi\hbar^2a\bar{n}/M \max(1/\omega_\perp,1/\omega_3)<\hbar$.
This condition is essentially the same as that for the LLL
approximation.
%Consequently, when the trap potential is removed, the particles
%fly apart rapidly, and do not have time to interact, becoming
%still more dilute as they expand.
%Therefore our approximation is
%consistent when the above criterion holds.
Further, if the interactions are enhanced by the use of a Feshbach
resonance (but still weak enough to use the LLL approx), then they
can be greatly reduced at the same time that the trap is turned
off, improving the accuracy of our neglect of interactions during
the expansion.

Finally, now that we have shown how the free expansion of a LLL
state acts as a wavefunction microscope, we point out that even a
single high-resolution snapshot taken after a free expansion of a
highly-correlated state contains a lot of information that can be
used as a diagnostic tool for the many-body physics. Such a
snapshot in principle gives a single typical configuration of all
the particle positions. In view of the preceding discussion, we
can discuss this in terms of the configuration at $t=0$, which is
drawn from the joint probability density
$|\psi_0(z_1,\ldots,z_N)|^2$ [we will assume the coordinates are
projected to the $1$--$2$ plane, so we drop $x_3$ and work in two
dimensions (2D) from here on]. In a highly correlated,
incompressible state such as Laughlin's, long-wavelength density
fluctuations are suppressed, and this can be seen even in a single
snapshot, if the particle number is large compared with 1. For
example, such a snapshot is shown in Ref.\ \cite{book}. It differs
markedly from a random configuration. The correlations can be
quantified by constructing the two-particle correlation function
of the snapshot. This is just a histogram of the values of ${\bf
r}_i -{\bf r}_j$ for all pairs of particles $i$, $j$ (it will be
automatically invariant under ${\bf r}\to {}-{\bf r}$). This can
be compared with the two-particle quantum- (and thermal-) average
correlation function $g(r)=\langle
\hat\psi^\dagger(z)\hat\psi^\dagger(z')\hat\psi(z')\hat\psi(z)\rangle/\bar{n}^2$
($\hat\psi(z)$ is the 2D field operator, and $r=|z-z'|$) which has
been calculated (usually in an edgeless geometry) for various
incompressible fluid ground states. We note that in the
thermodynamic limit, $g(r)$ is normalized so that it approaches
$1$ as $r\to\infty$. The wavefunction has an ergodicity property
that ensures that even a correlation function constructed from a
single sample (snapshot) reproduces the quantum/thermal-average
$g(r)$, provided the particle number is large. Further, the
Fourier transform of $\bar{n} g({\bf r})$ is essentially the
static (i.e.\ instantaneous) structure factor $s({\bf q})$
\cite{gmp}.

There are two basic results that can be extracted from the density
or its correlations $g(r)$ measured for a single snapshot. First,
if one examines a subregion (say, a square) of area $A$ of the
fluid and determines the particle number $N_s\leq N$ in this
region, this number will fluctuate as the subregion is moved over
a given snapshot, and also from one snapshot to another. If the
side of the square subregion is larger than the length $\xi$ we
define below, but small enough that the mean $\overline{N_s}$ is
$\ll N$, then in an equilibrium state the fluctuations $\Delta
N_s$ will be of order $(\Delta N_s)^2 = k_BT A\, d\bar{n}/d\mu$,
and give an estimate of the thermodynamic compressibility
$d\bar{n}/d\mu$ ($\mu$ is the chemical potential), if the
temperature $T$ is known. Alternatively, in the thermodynamic
limit one has $\lim_{{\bf q} \to 0}s({\bf
q})=(k_BT/\bar{n})d\bar{n}/d\mu$ \cite{forster}. In the
zero-temperature limit, $d\bar{n}/d\mu$ goes to a finite (either
zero or nonzero) value (except in the case of exactly-zero
interactions in a Bose gas!), so these fluctuations vanish. For a
translationally-invariant fluid in the LLL in the thermodynamic
limit, there is a LLL-projected version $\bar{s}({\bf q})$ of
$s({\bf q})$, which is easily obtained from the latter \cite{gmp}.
At $T=0$, $\bar{s}({\bf q})$ vanishes faster than ${\bf q}^2$ as
${\bf q}\to 0$, and for incompressible fluids (those in which
$\lim_{T\to0}d\bar{n}/d\mu=0$) it goes as ${\bf q}^4$ \cite{gmp}.
The latter behavior is analytic, implying that $g(r)-1$ tends to
zero rapidly at large $r$. At a finite temperature, there will be
a correlation length $\xi$, which diverges as $T\to 0$, and which
is defined by the property that $s({\bf q})$ will cross over at $q
\sim \xi^{-1}$ from the $T>0$ behavior at ${\bf q}\to0$ to the
$T=0$ behavior at larger $\bf q$. Note that in an incompressible
fluid such as Laughlin's, $\xi$ diverges as $\xi\sim e^{\Delta
E/(4k_BT)}$ as $T\to 0$, where $\Delta E $ is the
quasiparticle-hole excitation energy or ``gap''. The behavior at
$q>\xi^{-1}$ is the second property to look for using snapshots.
Thus from these correlation properties, it is possible in
principle to distinguish an incompressible quantum fluid from a
thermally-melted vortex lattice.

In practice, there will be both a resolution function convoluted
with the particle positions, and the question of the accuracy with
which the density is measured at any point (noise). However, our
proposal utilizes long-distance correlations where the spatial
resolution should not be a problem, while the average over
positions in a single snapshot alleviates the accuracy problem,
and reduces the noise, which is presumably uncorrelated.

We would like to contrast our remarks with a proposal of Sinova
{\it et al}.\ to measure the condensate fraction from the density
profile \cite{shm2}. They point out that, because of analyticity
properties in the LLL, the diagonal density matrix (i.e. the
quantum expectation $\langle n ({\bf r}) \rangle$ of the
single-particle density) determines the off-diagonal density
matrix from which the condensate fraction can be defined. Their
formula in our notation is
\begin{equation}
\beta=\frac{l_\perp^2}{4\pi N^2}\int d^2q\, |n({\bf q})|^2
e^{|{\bf q}|^2l_\perp^2/4},
\end{equation}
where $n({\bf q})$ is the Fourier transform of $\langle n ({\bf
r}) \rangle$. The normalization is such that the maximum possible
value of $\beta$ is $\beta=1$, and is attained in a product state,
as in eq.\ (\ref{bec}). With the help of examples \cite{shm2}, it
becomes clear that when the number of vortices $N_v$, say in a
vortex lattice state, is large, the integral is dominated by large
$\bf q$ values (up to $|\bf q|$ around $\sqrt{N_v}$), which tend
to be suppressed by multiplication by the Fourier transform of the
spatial resolution function. Further, even in a snapshot with
perfect resolution, the particle density would be a sum of
$\delta$-functions at the particle positions, and would differ
from the average density $\langle n ({\bf r}) \rangle$ because of
the presence of quantum (and more generally thermal) fluctuations.
These fluctuations are larger relative to the mean, $\langle
n({\bf r})\rangle$, at small values of $\nu$, which is precisely
the regime of greatest interest. They also remain large at large
$\bf q$ ($s({\bf q})\to 1$ \cite{gmp}), whereas $n({\bf q})\to 0$
\cite{shm2}, and (as for any noise in the determination of the
density at $\bf r$) they are enhanced in the integral by the
factor $e^{|{\bf q}|^2l_\perp^2/4}$. These fluctuations cannot be
removed by averaging over space without destroying the large $\bf
q$ information that is needed. It will be necessary to average
over many snapshots to obtain the quantum average $\langle n ({\bf
r}) \rangle$.

While we were completing this paper, a discussion with some
overlap with the second part of ours appeared \cite{demler}.

We thank I. Bloch, E.A. Cornell, M. Holland, J. Sinova, and
especially S.M. Girvin for useful discussions. We thank the
organizers and participants of the workshop on ``Correlation
Effects in Bose Condensates and Optical Lattices'' at the William
I. Fine Theoretical Physics Institute, University of Minnesota,
Minneapolis, for a stimulating environment in which this work was
done. We acknowledge support from NSF grant no.\ DMR-02-42949 (NR)
and EPSRC grant no.\ GR/R99027/01 (NRC).


\begin{references}

\vspace*{-15mm}


\bibitem{wgs}
% `do attractive bosons condense?'
N.K. Wilkin, J.M.F. Gunn and R.A. Smith, \prl {\bf 80}, 2265
(1998).

\bibitem{gp}
D.A. Butts and D.S. Rokhsar, Nature {\bf 397},  327  (1999); T.-L.
Ho, \prl {\bf 87}, 060403 (2001).

\bibitem{yrast}
B. Mottelson, Phys. Rev. Lett. {\bf 83},  2695  (1999); G.F.
Bertsch and T. Papenbrock, Phys. Rev. Lett. {\bf 83},  5412
(1999); A.D. Jackson, G.M. Kavoulakis, B. Mottelson, and S.M.
Reimann, Phys. Rev.  Lett. {\bf 86},  945  (2001); M.S. Hussein
and  O.K. Vorov, \pra {\bf 65}, 035603 (2002).

\bibitem{wgcw}
N.K. Wilkin and J.M.F. Gunn, \prl {\bf 84}, 6 (2000); N.R. Cooper
and N.K. Wilkin, \prb {\bf 60}, R16279 (1999).


\bibitem{cwg}
N.R. Cooper, N.K. Wilkin and J.M.F. Gunn, \prl {\bf 87}, 120405
(2001).


\bibitem{qh}
S. Viefers, T.H. Hansson and S.M. Reimann, Phys. Rev. A {\bf 62},
053604 (2000); B. Paredes, P. Fedichev, J.I. Cirac, and P. Zoller,
\prl {\bf 87}, 010402 (2001); J.W.~Reijnders, F.J.M.~van Lankvelt,
K.~Schoutens, and N.~Read, \prl {\bf 89}, 120401 (2002); B.
Paredes, P. Zoller, and J.I. Cirac, \pra {\bf 66}, 033609 (2002).

\bibitem{hm} T.-L. Ho and E.J. Mueller, \prl {\bf 89}, 050401
(2002).

\bibitem{shm1}
J. Sinova, C.B. Hanna, and A.H. MacDonald, \prl {\bf 89}, 030403
(2002).

\bibitem{shm2}
J. Sinova, C.B. Hanna, and A.H. MacDonald, \prl {\bf 90}, 120401
(2003).

\bibitem{kss}
Y. Kagan, E.L. Surkov, and G.V. Shlyapnikov, \pra
{\bf 54}, R1753 (1996).

\bibitem{cd1}
Y. Castin and R. Dum, \prl {\bf 77}, 5315 (1996).

\bibitem{cd2}
Y. Castin and R. Dum, Eur.\ Phys.\ J. D {\bf 7}, 399 (1999).

\bibitem{dm}
F. Dalfovo and M. Modugno, \pra {\bf 61}, 023605 (2000).

\bibitem{laugh}
R.B.~Laughlin, \prl {\bf 50}, 1395 (1983).

\bibitem{book}
R.B. Laughlin, Ch.\ 7 in
{\em The Quantum Hall Effect}, edited by R.E.~Prange and
S.M.~Girvin (Second Edition, Springer-Verlag, New York, 1990),
Fig.\ 7.7, p.\ 259.

\bibitem{gmp}
S.M.~Girvin, A.H.~MacDonald and P.~Platzman, \prb {\bf 33},
    2481 (1986).

\bibitem{forster} D. Forster, {\it Hydrodynamic Fluctuations,
Broken Symmetry, and Correlation Functions}, (Addison-Wesley,
Reading, MA, 1990), p.\ 27.

\bibitem{demler}
E. Altman, E. Demler, and M.D. Lukin, cond-mat/0306226.

\end{references}
\end{document}